\documentclass[12pt]{article}
\usepackage{amsmath}
\usepackage{graphicx}
\usepackage{subfigure}
\usepackage{hyperref}

\begin{document}

\author{Ernst Trojan \and \textit{Moscow Institute of Physics and Technology} \and 
\textit{PO Box 3, Moscow, 125080, Russia}}
\title{Dirac tachyons and antitachyons in many-particle system}
\maketitle

\begin{abstract}
A consistent description of charged many-tachyon Fermi system is developed.
Tachyons and antitachyons have the same chemical potential $\mu
_{+}=\mu _{-}$ because the axial coupling constant $g_{+}=g_{-}$ is
invariant under the charge conjugation, in contrast to reversion of the
electric charge $e_{+}=-e_{-} $. The axial density $n_5=\left\langle \bar
\psi \gamma ^0\gamma _5\psi \right\rangle $ is incorporated in the
thermodynamical functions instead of $\left\langle \bar \psi \gamma ^0\psi
\right\rangle$ which is not associated with any conserved quantity. The
number of tachyons and antitachyons are undefined but it is possible to
estimate their difference and establish a link between the total electric
charge density $en $ and $n_5$.
\end{abstract}

\section{Introduction}

Tachyon is a substance that moves faster than light. Its energy spectrum
satisfies dispersion relation 
\begin{equation}
\varepsilon _p^2=p^2-m^2  \label{tah}
\end{equation}
and its group velocity 
\begin{equation}
v=\frac{d\varepsilon _p}{dp}=\frac p{\sqrt{p^2-m^2}}  \label{tah1}
\end{equation}
exceeds the speed of light in vacuum $c=1$, relative to any reference frame.

Tachyons are commonly known in the field theory \cite{C1} and nonlinear
optics \cite{O1}. The Lagrangian of a free fermionic tachyon \cite{D1,D2}$%
L_0=\bar \psi \left( i\gamma _5\gamma ^\mu \partial _\mu -m\right) \psi $
corresponds to the equation of motion 
\begin{equation}
\left( i\gamma ^\mu \partial _\mu -\gamma _5m\right) \psi =0  \label{d0}
\end{equation}
whose plane wave solution 
\begin{equation}
\psi =\left( 
\begin{array}{c}
\phi \\ 
\chi
\end{array}
\right) \exp \left( i\vec p\cdot \vec r-i\varepsilon _{p\,}t\right)
\label{wa}
\end{equation}
results in the single-particle energy spectrum (\ref{tah}). The motion of
tachyon in the presence of electromagnetic field $A_\mu $ is described by
equation \cite{D3} 
\begin{equation}
\left( i\gamma ^\mu \left( \partial _\mu +ieA_\mu \right) -\gamma _5m\right)
\psi =0  \label{d2}
\end{equation}
that corresponds to appearance of interaction term $-e\bar \psi \gamma
_5\gamma ^\mu A_\mu \psi $ in the free tachyonic Lagrangian. The equation of
motion 
\begin{equation}
\left( \gamma ^\mu \left( i\partial _\mu -eA_\mu \right) -\gamma _5m-g\gamma
_5\gamma ^\mu \omega _\mu \right) \psi =0  \label{d3}
\end{equation}
is written when the relevant Lagrangian $L=L_0-e\bar \psi \gamma _5\gamma
^\mu A_\mu \psi -g\bar \psi \gamma ^\mu \omega _\mu \psi $ includes the
coupling term associated with vector field $\omega _\mu $.

When we consider an ensemble of particle and antiparticles in finite volume $%
V=\int d^3r$, we use standard definitions of the particle number density 
\begin{equation}
n\equiv \left\langle \bar \psi \gamma ^0\psi \right\rangle =\frac \gamma
{\left( 2\pi \right) ^3}\int \bar \psi \gamma ^0\psi \,d^3p  \label{nn0}
\end{equation}
and axial density 
\begin{equation}
n_5\equiv \left\langle \bar \psi \gamma ^0\gamma _5\psi \right\rangle =\frac
\gamma {\left( 2\pi \right) ^3}\int \bar \psi \gamma ^0\gamma _5\psi \,d^3p
\label{n50}
\end{equation}
The relevant total charges 
\begin{equation}
\left\langle Q_e\right\rangle \equiv \left( e_{+}n_{+}+e_{-}n_{-}\right) V
\label{ee}
\end{equation}
\begin{equation}
\left\langle Q_g\right\rangle \equiv \left( g_{+}n_{5+}+g_{-}n_{5-}\right) V
\label{gg}
\end{equation}
include contributions from the two species that depend on the sign of
elementary charges of particles $e_{+}$, $g_{+}$ and antiparticles $e_{-}$, $%
g_{-}$.

The electric charge conjugation \cite{D3} $e_{+}=-e_{-}\equiv e$ implies
that the total electric charge of tachyons and antitachyons 
\begin{equation}
Q_e\equiv e\left( n_{+}-n_{-}\right) V  \label{ee2}
\end{equation}
coincides with the relevant expression for ordinary fermions and
antifermions \cite{KaZ}. Most natural assumption $g_{-}=-g_{+}$ would lead
to formula similar to (\ref{ee2}) but relation between $g_{+}$ and $g_{-}$
is not evident beforehand. It is necessary to check the charge conjugation
of tachyonic Dirac equation (\ref{d3}) and establish what consequences it
implies to the thermodynamics of a many-tachyon system.

\section{Charge conjugation}

Equation (\ref{d3}) is equivalent to 
\begin{equation}
\left( \gamma _5\gamma ^\mu \left( i\partial _\mu -eA_\mu \right) -m-g\gamma
^\mu \omega _\mu \right) \psi =0  \label{d3b}
\end{equation} 
The difference between $e$ and $g$ is evident from the corresponding bilinear transforms.
Let us demonstrate it directly, following the previous analysis \cite{D3} and considering the charge conjugation of equation (\ref{d3}) or (\ref{d3b}). 
The action of Hermite operator $\dagger $
(transposition and complex conjugation) results in
\begin{equation}
\psi ^{\dagger }\left( -i\left[ \gamma _5\gamma ^\mu \right] ^{\dagger
}\partial _\mu -e\left[ \gamma _5\gamma ^\mu \right] ^{\dagger }A_\mu
-m-g\gamma ^{\mu \dagger }\omega _\mu \right) =0  \label{d2a}
\end{equation}
Presenting $\psi ^{\dagger }\equiv \psi ^{\dagger }\gamma ^0\gamma ^0=\bar
\psi \gamma ^0$ and multiplying equation (\ref{d2a}) by $\gamma ^0$ from the
right, we have 
\begin{equation}
\bar \psi \gamma ^0\left( -i\left[ \gamma _5\gamma ^\mu \right] ^{\dagger
}\partial _\mu -e\left[ \gamma _5\gamma ^\mu \right] ^{\dagger }A_\mu
-m-g\gamma ^{\mu \dagger }\omega _\mu \right) \times \gamma ^0=0  \label{d2c}
\end{equation}
Using identities 
\begin{equation}
\gamma ^0\gamma ^{\mu \dagger }\gamma ^0=\gamma ^\mu \qquad \gamma ^\mu
\gamma _5=-\gamma _5\gamma ^\mu \qquad \gamma _5=\gamma _5^{\dagger }\qquad
\gamma ^0\gamma _5\gamma ^0=-\gamma _5  \label{dma}
\end{equation}
and 
\begin{equation}
\gamma ^0\left[ \gamma _5\gamma ^\mu \right] ^{\dagger }\gamma ^0=\gamma
^0\gamma ^{\mu \dagger }\gamma _5^{\dagger }\gamma ^0=\gamma ^0\gamma ^{\mu
\dagger }\gamma ^0\gamma ^0\gamma _5\gamma ^0=\gamma _5\gamma ^\mu
\label{dma2}
\end{equation}
we obtain 
\begin{equation}
\bar \psi \left( -i\gamma _5\gamma ^\mu \partial _\mu -e\gamma _5\gamma ^\mu
A_\mu -m-g\gamma ^\mu \omega _\mu \right) \gamma ^0=0  \label{d2d}
\end{equation}
Transposition of (\ref{d2d}) yields 
\begin{equation}
\left( i\left[ \gamma _5\gamma ^\mu \right] ^T\partial _\mu +e\left[ \gamma
_5\gamma ^\mu \right] ^TA_\mu +m+g\gamma ^{\mu T}\omega _\mu \right) \bar
\psi ^T=0  \label{d2e}
\end{equation}
that can be also written in the equivalent form 
\begin{equation}
\left( i\gamma ^{\mu T}\partial _\mu +e\gamma ^{\mu T}A_\mu -\gamma
_5m+g\left[ \gamma _5\gamma ^\mu \right] ^T\omega _\mu \right) \bar \psi ^T=0
\label{d2f}
\end{equation}

Let us introduce the charge conjugation matrix $C$ with properties 
\begin{equation}
C\gamma ^{\mu T}C^{-1}=-\gamma ^\mu  \label{cc}
\end{equation}
that performs reversal 
\begin{equation}
Ce\gamma ^{\mu T}A_\mu C^{-1}=-e\gamma ^\mu A_\mu \qquad Cg\gamma ^{\mu
T}\omega _\mu C^{-1}=-g\gamma ^\mu \omega _\mu  \label{ce}
\end{equation}
Identity (\ref{cc}) immediately implies 
\begin{equation}
C\left[ \gamma _5\gamma ^\mu \right] ^TC^{-1}=C\gamma ^{\mu T}CC^{-1}\gamma
_5C^{-1}=-\gamma ^\mu \gamma _5=\gamma _5\gamma ^\mu  \label{ce1}
\end{equation}
and 
\begin{eqnarray}
&C\gamma _5C^{-1} =iC\left[ \gamma ^0\gamma ^1\gamma ^2\gamma ^3\right]
^TC^{-1}=iC\gamma ^{3T}\gamma ^{2T}\gamma ^{1T}\gamma ^{0T}C^{-1} =& 
\nonumber \\
&=i\left(C\gamma ^{3T}C\right) \left( C^{-1}\gamma ^{2T}C\right) \left(
C^{-1}\gamma ^{1T}C\right) \left( C\gamma ^{0T}C^{-1}\right) = &  \nonumber
\\
&=i\left( -\gamma ^3\right) \left( -\gamma ^2\right) \left( -\gamma
^1\right) \left( -\gamma ^0\right) =i\gamma ^3\gamma ^2\gamma ^1\gamma
^0=\gamma _5&  \label{dir}
\end{eqnarray}
Then, multiplying equation (\ref{d2f}) by $C$ from the left, we have 
\begin{equation}
C\times \left( i\gamma ^{\mu T}\partial _\mu +e\gamma ^{\mu T}A_\mu -\gamma
_5m+g\left[ \gamma _5\gamma ^\mu \right] ^T\omega _\mu \right) C^{-1}C\bar
\psi ^T=0  \label{d2g}
\end{equation}
hence, 
\begin{equation}
\left( i\gamma ^\mu \partial _\mu +e\gamma ^\mu A_\mu +\gamma _5m-g\gamma
_5\gamma ^\mu \omega _\mu \right) \bar \psi ^C=0  \label{d2h}
\end{equation}
where $\psi ^C=C\bar \psi ^T$ is the charge conjugated wave function.
Alternatively, multiplying equation (\ref{d2e}) by $C$ from the left, we
have 
\begin{equation}
C\times \left( i\left[ \gamma _5\gamma ^\mu \right] ^T\partial _\mu +e\left[
\gamma _5\gamma ^\mu \right] ^TA_\mu +m+g\gamma ^{\mu T}\omega _\mu \right)
C^{-1}C\bar \psi ^T=0  \label{d2i}
\end{equation}
hence 
\begin{equation}
\left( i\gamma _5\gamma ^\mu \partial _\mu +e\gamma _5\gamma ^\mu A_\mu
+m-g\gamma ^\mu \omega _\mu \right) \psi ^C=0  \label{d2j}
\end{equation}
that is equivalent to (\ref{d2h}).

It is not a problem that the tachyonic Dirac equation (\ref{d3}) is not
invariant under the charge conjugation on account of wrong sign of the
mass term in (\ref{d2h}) because equation (\ref{d3}) is still CP and T invariant 
\cite{D3}. The important fact we have seen now is that the charge
conjugation implies no more than electric charge reversion $e\leftrightarrow
-e$ and does not concern the axial coupling constant $g$. 
Thus, tachyons and antitachyons have opposite $e_{+}=-e_{-}$ but the same $g_{+}=g_{-}$. Hence, in contrast to the total electric charge (\ref{ee2}),
the total axial charge (\ref{gg}) is calculated by formula 
\begin{equation}
\left\langle Q_g\right\rangle \equiv g\left( n_{5+}+n_{5-}\right) V=\mathrm{%
const}  \label{ggg}
\end{equation}

\section{Thermodynamical functions}

According to the Noether theorem, a conserved current corresponds to each
continuous symmetry of the Lagrangian. The relevant partition function is
constructed so that its argument $L+\mu O$ includes the chemical potential $%
\mu $ as a free multiplier and the relevant conserved quantity $O$ \cite{KaZ}%
. The tachyonic equation of motion (\ref{d0}) results in the continuity
equation \cite{T47,T89} $\partial _{\mu \,}j_5^\mu =0$ of the axial current $%
j_5^\mu =(j_5^0,\vec j_5)=\bar \psi \gamma ^\mu \gamma _5\psi $. Integration 
\begin{equation}
\int \left( \partial _{0\,}j_5^0\,+\mathrm{div\,}\vec j_5\right) d^3r=\int
\partial _{0\,}j_5^0\,\,d^3r  \label{cnt}
\end{equation}
implies conservation of quantity 
\begin{equation}
N_5=\int j_5^0d^3r=\int \bar \psi \gamma ^0\gamma _5\psi \,d^3r  \label{q5}
\end{equation}
called as the axial charge or axial particle number. The vector current $%
j^\mu =\bar \psi \gamma ^\mu \psi $ obeys equation \cite{T89} 
\begin{equation}
\partial _{\mu \,}j^\mu =-2im\bar \psi \gamma _5\psi  \label{eqn}
\end{equation}
and quantity 
\begin{equation}
N=\int j^0d^3r=\int \bar \psi \gamma ^0\psi \,d^3r  \label{N}
\end{equation}
is not conserved (except the only chiral limit $m\rightarrow 0$). Therefore,
it is the axial charge (\ref{q5}) which is incorporated in the partition
function of free tachyon Fermi gas 
\begin{equation}
Z=\int \left[ d\bar \psi _{+}\right] \left[ d\psi _{-}\right] \left[ d\bar
\psi _{-}\right] \left[ d\psi _{-}\right] \exp \left\{
\int\limits_0^{1/T}d\tau \left( \int L_{+}d^3r+\int L_{-}d^3r+\mu
_{+}N_{5+}+\mu _{-}N_{5-}\right) \right\}  \label{z1}
\end{equation}
where $\tau =it$. We emphasize that the number of particles (\ref{N}) is not
included in the partition function but the axial charge (\ref{q5})
is the main characteristic of tachyon Fermi gas. Replacement $%
N\leftrightarrow N_5$ brings no formal change to the thermodynamical laws, however,
the two quantities should not be mixed when we deal with
the relevant charges (\ref{ee2}) and (\ref{ggg}): the quantity associated with the
total electric charge is not conserved.

The tachyonic Dirac equation is also time invariant \cite{D3}.\textrm{\ }If
time reversal is a good symmetry, a detailed balance must occur among all
possible reactions in equilibrium and the Gibbs free energy will remain
constant
\begin{equation}
G=\mu _{+}n_{5+}V+\mu _{-}n_{5-}V=\mathrm{const}  \label{gib}
\end{equation}
In the light of (\ref{ggg}) it implies that tachyons and antitachyons must
have the same chemical potential 
\begin{equation}
\mu \equiv \mu _{+}=\mu _{-}  \label{mu}
\end{equation}
This allows to simplify (\ref{z1}) in the form$\ Z=Z_{\pm }^2$ where 
\begin{equation}
Z_{\pm }\equiv \int \left[ d\bar \psi \right] \left[ d\psi \right] \exp
\left\{ \int\limits_0^{1/T}d\tau \left( \int Ld^3r+\mu N_5\right) \right\}
\label{z2}
\end{equation}
Then, the pressure, energy density and entropy are determined by standard
formulas 
\begin{equation}
P=\frac TV\ln Z=\frac \gamma {2\pi ^2}T\int\limits_0^\infty \ln \left(
1+\exp \frac{\mu -\varepsilon _p}T\right) p^2dp=\frac \gamma {6\pi
^2}\int\limits_0^\infty \,f_\varepsilon \,\frac{d\varepsilon _p}{dp}p^2dp
\label{p0}
\end{equation}
\begin{equation}
E=\frac{T^2}V\left( \frac{\partial \ln Z}{\partial T}\right) _{V,\mu }+\mu
n_5=\frac \gamma {2\pi ^2}\int\limits_0^\infty \,f_\varepsilon \,\varepsilon
_pp^2dp  \label{e}
\end{equation}
\begin{equation}
S=V\frac{E+P-\mu n_5}T  \label{s}
\end{equation}
where 
\begin{equation}
f_\varepsilon =\frac 1{\exp \left[ \left( \varepsilon _p-\mu \right)
/T\right] +1}  \label{f}
\end{equation}
is the Fermi-Dirac distribution function, and the axial density satisfies
identity

\begin{equation}
n_5\equiv T\left( \frac{\partial \ln Z}{\partial \mu }\right) _{V,T}=\frac
\gamma {2\pi ^2}\int\limits_0^\infty f_\varepsilon \,p^2dp  \label{n5}
\end{equation}
Since the thermodynamical functions of free tachyons and antitachyons are
indistinguishable, the proper degeneracy factor of tachyons $\gamma =1$ is
doubled so that $\gamma =2$ in all thermodynamical formulas (or we can write 
$2\gamma $ implying that $\gamma =1$).

Note that the thermodynamical functions of ordinary baryonic matter are given by
formulas \cite{KaZ} 
\begin{equation}
P=\frac \gamma {6\pi ^2}\int\limits_0^\infty \,\left( f_{\varepsilon
+}+f_{\varepsilon -}\right) \,\frac{d\varepsilon _p}{dp}p^2dp  \label{p0-}
\end{equation}
\begin{equation}
E=\frac \gamma {2\pi ^2}\int\limits_0^\infty \,\left( f_{\varepsilon
+}+f_{\varepsilon -}\right) \,\varepsilon _pp^2dp  \label{e-}
\end{equation}
\begin{equation}
S=V\frac{E+P-\mu n}T  \label{s--}
\end{equation}
where
\begin{equation}
n=\frac \gamma {2\pi ^2}\int\limits_0^\infty \,\left( f_{\varepsilon
+}-f_{\varepsilon -}\right) \,p^2dp  \label{n5-}
\end{equation}
is the baryon number density, while distribution functions of particles and
antiparticles are 
\begin{equation}
f_{\varepsilon +}=\frac 1{\exp \left[ \left( \varepsilon _p-\mu \right)
/T\right] +1}\qquad f_{\varepsilon -}=\frac 1{\exp \left[ \left( \varepsilon
_p+\mu \right) /T\right] +1}  \label{f-}
\end{equation}
Indeed, the tachyonic formulas (\ref{p0})-(\ref{n5}) do immediately follow
from the relevant formulas of hot baryonic matter (\ref{p0-})-(\ref{f-}) if
particles and antiparticles have the same chemical potential and $%
f_{\varepsilon +}=f_{\varepsilon -}$.

For a cold tachyon Fermi gas with Fermi momentum 
\begin{equation}
p_F>m  \label{pf}
\end{equation}
the distribution function (\ref{f}) is reduced to the Heaviside step 
\begin{equation}
f_\varepsilon =\Theta \left( \varepsilon_p -\varepsilon _F\right) =\Theta
\left( p-p_F\right)  \label{hea}
\end{equation}
where\thinspace
\begin{equation}
\varepsilon _F=\sqrt{p_F^2-m^2}  \label{zer}
\end{equation}
is the tachyon Fermi energy. The axial charge density (\ref{n5}) is
immediately calculated 
\begin{equation}
n_5=\frac{\gamma p_F^3}{6\pi ^2}  \label{n55}
\end{equation}
while formula (\ref{s}) is reduced to 
\begin{equation}
E+P=\mu n_5  \label{rd3}
\end{equation}
instead of ordinary $E+P=\mu n$. As we have emphasized above, the axial charge density $n_5$ appears in all thermodynamical formulas instead of the particle number density of the
ordinary Fermi gas. Formula (\ref{n55}) is\ valid under condition (\ref{pf})
when the axial charge density exceeds 
\begin{equation}
n_5>n_{\star }=\frac{\gamma m^3}{6\pi ^2}  \label{nt}
\end{equation}
Of course, it does not imply that the number of tachyons must also exceed some finite bottom level. For thermodynamical relations (\ref{z1}), (\ref{p0})-(\ref{n5})
does not provide us any information about quantity 
\begin{equation}
\left\langle \bar \psi \gamma ^0\psi \right\rangle \,  \label{psi}
\end{equation}
and it is not clear whether it is finite of has any physical meaning because
the number of tachyons (\ref{N}) is not conserved.

\section{Scalar and particle number density}

Substituting plane-wave solution (\ref{wa}) in the tachyonic equation (\ref
{d0}), we get a linear system for bispinors 
\begin{equation}
\begin{array}{c}
\left( \vec \sigma \cdot \vec p-m\right) \phi =\varepsilon _p\chi \\ 
\left( \vec \sigma \cdot \vec p+m\right) \chi =\varepsilon _p\phi
\end{array}
\label{bi}
\end{equation}
Hence, 
\begin{equation}
\chi =\frac{\varepsilon _p}{hp+m}\phi =\frac{hp-m}{\varepsilon _p}\phi
\label{bi2}
\end{equation}
where 
\begin{equation}
h=\frac{\vec \sigma \cdot \vec p}p=\pm 1  \label{hel}
\end{equation}
is helicity of tachyon (or antitachyon). It is clear that the sign of
helicity remains the same regardless of the point of view of external observer moving at arbitrary subluminal velocity. Hence, if $h=+1$, say for tachyons, it is always $%
h=-1$ for antitachyons and v.v.

In the light of bispinor representation (\ref{wa}), we define the following
quantities 
\begin{equation}
j^0=\bar \psi \gamma ^0\psi =\left\| \phi \right\| ^2+\left\| \chi \right\|
^2=\left[ 1+\frac{\left| \varepsilon _p\right| ^2}{\left( hp+m\right) ^2}%
\right] \left\| \phi \right\| ^2=\left\{ 
\begin{array}{cc}
\cfrac{2hp\left\| \phi \right\| ^2 }{hp+m} & p\geq m \\ 
\cfrac{2m\left\| \phi \right\| ^2 }{hp+m} & p<m
\end{array}
\right.  \label{jj0}
\end{equation}
\begin{equation}
j_s=\bar \psi \psi =\left\| \phi \right\| ^2-\left\| \chi \right\| ^2=\left[
1-\frac{\left| \varepsilon _p\right| ^2}{\left( hp+m\right) ^2}\right]
\left\| \phi \right\| ^2=\left\{ 
\begin{array}{cc}
\cfrac{2m\left\| \phi \right\| ^2 }{hp+m} & p\geq m \\ 
\cfrac{2hp\left\| \phi \right\| ^2 }{hp+m} & p<m
\end{array}
\right.  \label{jjs}
\end{equation}
\begin{equation}
j_5^0=\bar \psi \gamma ^0\gamma _5\psi =\phi ^{*}\chi +\chi ^{*}\phi =\frac{2%
\mathrm{Re}\varepsilon _p}{hp+m}\left\| \phi \right\| ^2  \label{jj05}
\end{equation}

The axial charge density $n_5=\left\langle j_5^0\right\rangle $ is
determined by formula (\ref{n5}). Taking also into account formulas (\ref{nn0}), (\ref{n50}), (\ref{jj0}) and (\ref{jj05}%
), we find general expression for the particle number density 
\begin{equation}
\left\langle j^0\right\rangle \equiv \left\langle \bar \psi \gamma ^0\psi
\right\rangle =\frac \gamma {2\pi ^2}\int\limits_m^\infty f_\varepsilon
\frac h{\left| \varepsilon _p\right| }p^3dp+\frac \gamma {2\pi
^2}\int\limits_0^m\frac m{\mathrm{Re}\varepsilon _p}\mathrm{\ }p^2dp
\label{xn}
\end{equation}
and, in the light of (\ref{jjs}) and (\ref{jj05}) we find general expression
for the scalar density 
\begin{equation}
\left\langle j_s\right\rangle \equiv \left\langle \bar \psi \psi
\right\rangle =\frac \gamma {2\pi ^2}\int\limits_m^\infty f_\varepsilon
\frac m{\left| \varepsilon _p\right| }p^2dp+\frac \gamma {2\pi
^2}\int\limits_0^mf_{\varepsilon \,}\frac{\mathrm{\ }h}{\mathrm{Re}%
\varepsilon _p}p^3dp  \label{xs}
\end{equation}

Since the particles and antiparticles have opposite helicity, we immediately
write expressions each number density  
\begin{equation}
n_{+}=\left\langle j_{+}^0\right\rangle =\frac \gamma {2\pi
^2}\int\limits_m^\infty f_{\varepsilon +}\frac h{\left| \varepsilon
_p\right| }p^3dp+\frac \gamma {2\pi ^2}\int\limits_0^mf_{\varepsilon -}\frac
m{\mathrm{Re}\varepsilon _p}\mathrm{\ }p^2dp  \label{n+}
\end{equation}
\begin{equation}
n_{-}=\left\langle j_{-}^0\right\rangle =\frac \gamma {2\pi
^2}\int\limits_m^\infty f_{\varepsilon +}\frac{\left( -h\right) }{\left|
\varepsilon _p\right| }p^3dp+\frac \gamma {2\pi
^2}\int\limits_0^mf_{\varepsilon -}\frac m{\mathrm{Re}\varepsilon _p}\mathrm{%
\ }p^2dp\,\,  \label{n-}
\end{equation}
and for the scalar density of particles and antiparticles 
\begin{equation}
\left\langle j_{s+}\right\rangle =\frac \gamma {2\pi ^2}\int\limits_m^\infty
f_{\varepsilon +}\frac m{\left| \varepsilon _p\right| }p^2dp+\frac \gamma
{2\pi ^2}\int\limits_0^mf_{\varepsilon -}\frac{\mathrm{\ }h}{\mathrm{Re}%
\varepsilon _p}p^3dp  \label{s+}
\end{equation}
\begin{equation}
\left\langle j_{s-}\right\rangle =\frac \gamma {2\pi ^2}\int\limits_m^\infty
f_{\varepsilon +}\frac m{\left| \varepsilon _p\right| }p^2dp+\frac \gamma
{2\pi ^2}\int\limits_0^mf_{\varepsilon -}\frac{\mathrm{\ }\left( -h\right) }{%
\mathrm{Re}\varepsilon _p}p^3dp  \label{s-}
\end{equation}
Taken into account that $n_{5+}=n_{5-}$ and $f_{\varepsilon
+}=f_{\varepsilon -}\equiv f_\varepsilon $ because particles and
antiparticles have the same chemical potential (\ref{mu}), we find by means
of (\ref{ee2}) the total electric charge 
\begin{eqnarray}
&Q_e =\left( e_{+}n_{+}+e_{-}n_{-}\right) V=&e\left\langle \left( \frac{hp}{%
\left| \varepsilon _p\right| }+\frac m{\mathrm{Re}\varepsilon _p}\mathrm{\ }%
\right) j_5^0\right\rangle V-e\left\langle \left( \frac{-hp}{\left|
\varepsilon _p\right| }+\frac m{\mathrm{Re}\varepsilon _p}\mathrm{\ }\right)
j_5^0\right\rangle V=  \nonumber \\
&&=\frac{\gamma Veh}{2\pi ^2}\int\limits_m^\infty f_\varepsilon \frac{p^3dp}{%
\sqrt{p^2-m^2}} \quad \quad \quad  \label{q2}
\end{eqnarray}
and determine 
\begin{equation}
n=n_{+}-n_{-}=h\tilde n=\frac{\gamma h}{2\pi ^2}\int\limits_m^\infty
f_\varepsilon \frac{p^3dp}{\sqrt{p^2-m^2}}\,\,  \label{n2}
\end{equation}
as the tachyonic ''particle number density''. It is finite in spite of the
fact that each contribution of tachyons (\ref{n+}) and antitachyons (\ref{n-}%
) are separately divergent. We deliberately leave helicity $h$ in (\ref{n2})
because it implies correct definition of $n\equiv \left\langle \bar \psi
\gamma ^0\psi \right\rangle _{+}-\left\langle \bar \psi \gamma ^0\psi
\right\rangle _{-}$. The total electric charge $Q_e=en=eh\tilde n$ depends
on the handedness and the sign of tachyonic charge: for example, if
left-handed tachyon ($h=-1$) has electric charge $e$ equal to 1 electron
charge, then, the total electric charge of a many-tachyon system $Q_e$ has
always opposite sign to the charge of electron. This can be compared with a
hot nuclear matter where the number of protons always exceeds the number of
antiprotons so that the total electric charge of nucleon is always positive
(and counterbalanced by the negative charge of electron gas). Quantity 
\begin{equation}
\tilde n=\frac \gamma {2\pi ^2}\int\limits_m^\infty f_\varepsilon \frac{p^3dp%
}{\sqrt{p^2-m^2}}\,\,  \label{nm}
\end{equation}
is always positive and it can be called as effective particle number
density, bearing in mind how it is incorporated in (\ref{q2})-(\ref{n2}). At
zero temperature (\ref{hea}) it is reduced to 
\begin{equation}
\tilde n=\left| n\right| =\frac \gamma {6\pi ^2}\varepsilon _F\left(
\varepsilon _F^2+3m^2\right) \,  \label{n22}
\end{equation}
and the total electric charge (\ref{q2}) is easily expressed 
\begin{equation}
Q_e=\frac{eh\gamma V}{6\pi ^2}\sqrt{p_F^2-m^2}\left( p_F^2+2m^2\right) \,=%
\frac{eh\gamma V}{6\pi ^2}\sqrt{\left( \frac{6\pi ^2n_5}\gamma \right)
^{2/3}-m^2}\left[ \left( \frac{6\pi ^2n_5}\gamma \right) ^{2/3}+2m^2\right]
\,  \label{q22}
\end{equation}
in terms of the axial density $n_5$ (\ref{n55}).

The total scalar density 
\begin{equation}
n_s=\left\langle j_{s+}\right\rangle +\left\langle j_{s-}\right\rangle
=\left\langle \left( \frac m{\left| \varepsilon _p\right| }+\frac{hp}{%
\mathrm{Re}\varepsilon _p}\right) j_5^0\right\rangle +\left\langle \left(
\frac m{\left| \varepsilon _p\right| }+\frac{-hp}{\mathrm{Re}\varepsilon _p}%
\right) j_5^0\right\rangle =\frac{\gamma m}{2\pi ^2}\int\limits_m^\infty
f_\varepsilon \frac{p^2dp}{\sqrt{p^2-m^2}}\,  \label{ns}
\end{equation}
is also finite while each contribution of particles (\ref{s+}) and
antiparticles (\ref{s-}) are divergent. The scalar density (\ref{ns}) at
zero temperature is determined by formula 
\begin{equation}
n_s=\frac{\gamma m}{4\pi ^2}\left( p_{F\,}\varepsilon _F+m^2\ln \frac{%
p_F+\varepsilon _F}m\right)  \label{ns22}
\end{equation}
that coincides with expression derived in the earlier work \cite{TV2011c} in
by means of intuitive analysis but without strict proof. Note that the
scalar density of an ordinary fermionic matter 
\begin{equation}
n_s=\frac{\gamma m}{4\pi ^2}\left( p_{F\,}\varepsilon _F-m^2\ln \frac{%
p_F+\varepsilon _F}m\right)  \label{ns44}
\end{equation}
differs from (\ref{ns22}) by the sign before $m^2$.

The knowledge of (\ref{n5}), (\ref{n2}) and (\ref{ns}) is necessary for calculation the of thermodynamical functions of interacting tachyon Fermi gas.
Comparing formulas (\ref{n55}), (\ref{n22}) and (\ref{ns22}), one notes that 
$n>n_s$ at any $p_F$, and $n_s>n_5$ when $1.08m<p_F<1.80m$, while $n>n_5$
when $p_F>1.06m$, see Fig.~\ref{ns5}. Maximum ratio $n/n_5\simeq 1.41$ is
achieved at $p_F\simeq 1.41$. At large $p_F\gg m$ and $\varepsilon
_F\rightarrow p_F$ the tachyon matter behaves as an ordinary massless Fermi
gas, and $n\rightarrow n_5$. The limit of low density $n\rightarrow 0$
corresponds to $\varepsilon _F\rightarrow 0$ and $p_F\rightarrow m$ while
the minimum axial density (\ref{nt}) is achieved at the vanishing particle number density $n=0$.
However, this limit is not achieved in practice because a tachyon Fermi gas is
unstable with respect to hydrodynamical perturbations at such small density
since their causal propagation takes place only at \cite{TV2011c} 
\begin{equation}
p_F\geq \sqrt{\frac 32}m\,  \label{pt}
\end{equation}
that, in the light of (\ref{n22}), corresponds to 
\begin{equation}
n\geq n_c=\frac{5\sqrt{2}\gamma }{24\pi ^2}m^3\,  \label{nc}
\end{equation}
The tachyon medium can exist only at finite material density 
\begin{equation}
\rho \geq mn_c=\frac{\gamma 5\sqrt{2}}{24\pi ^2}m^4\,  \label{r}
\end{equation}
while a rarefied tachyon Fermi gas will be unstable, perhaps, decaying
into an aggregate of dense droplets.

Of course, we could avoid divergences in (\ref{n+})-(\ref{s-}), considering
the only sector $p>m$ and excluding the low-momentum states $p<m$ as
unphysical ones. However, in spite of attractiveness of this approach, it
results in serious contradictions so that statistical description of a
many-tachyon system becomes senseless \cite{T2012b}.

\section{Conclusion}

A many-tachyon system is looking strange and contrasting to an ordinary
system of particles and antiparticles. The main invariant is the axial
charge (\ref{q5}) while the number of tachyons (\ref{N}) is not conserved
(like the number of photons in black body radiation, or the number of thermal
excitation in solids). The axial charge density $n_5$ determined by formula (%
\ref{n5}) is incorporated in the thermodynamical equations (\ref{z2})-(\ref
{n5}) of a tachyon Fermi gas instead of the particle number density $n$ (\ref{nn0}%
).

The charge conjugation (\ref{d2j}) changes the signs of electric charge
while the axial charge remains the same implying that the tachyons and
antitachyons have equal chemical potential $\mu _{+}=\mu _{-}$ (\ref{mu}).
This fact is crucial in the thermodynamics of tachyon Fermi gas because
negative $\mu _{-}=-\mu _{+}<0$ would not allow us to operate with unambiguous
distribution function (\ref{f}) at small momentum $p<m$, implying impossibility 
of regularization of the total electric charge (\ref{q2}). 

\begin{figure}[tbp]
\caption{Axial (solid), particle number (dashed) and scalar (dotted)
densities of tachyon Fermi gas at zero temperature vs Fermi momentum}
\label{ns5}{\includegraphics[scale=0.8]{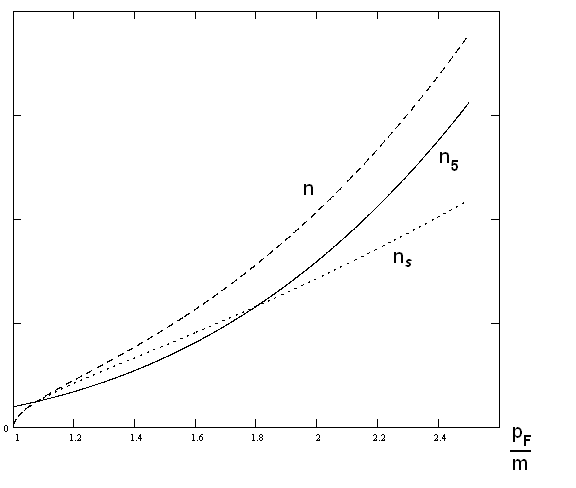}}
\end{figure}

The number of tachyons $N_{+}$ and antitachyons $N_{-}$ as well as their
summary number $N_{+}+N_{-}$ cannot be defined or presented as a function on
temperature. Nevertheless, the difference between the particles and
antiparticles $N_{+}-N_{-}$ is reflected in the total electric charge (\ref
{ee2}) and we managed to estimate it in terms of the thermodynamical
functions of tachyon gas (\ref{q2}). The scalar density is also estimated (%
\ref{ns}) and at zero temperature it coincides with expression found in the
previous analysis \cite{TV2011c}. At zero temperature the axial density,
particle number density, total electric charge and scalar density are given by formulas (\ref
{n55}), (\ref{n22}), (\ref{q22}) and (\ref{ns22}), respectively, see Fig.~%
\ref{ns5}.

As for the alternative Dirac equation with imaginary mass 
\begin{equation}
\left( i\gamma ^\mu \partial _\mu -im\right) \psi =0  \label{dii}
\end{equation}
it yields the same tachyonic dispersion relation (\ref{tah}). However, it is
not associated with any conserved current because \cite{T47,T89} $\partial
_{\mu \,}j^\mu =2m\bar \psi \psi \neq 0$ and $\partial _{\mu \,}j_5^\mu
\rightarrow \infty $. The relevant partition function will be equivalent to (%
\ref{z1}) at zero chemical potential, i.e. when tachyonic thermal
excitations are considered.

When a may-tachyon system is put in external filed, the single-particle
energy spectrum of particles $\varepsilon _{p+}$ and antiparticles $%
\varepsilon _{p-}$ will be different. The thermodynamical functions of the
whole system will be calculated by formulas (\ref{p0-})-(\ref{n5-}) and the
distribution functions of the tachyons $f_{\varepsilon +}$ and antitachyons $%
f_{\varepsilon -}$ may not coincide. Then, the total electric charge (\ref
{n2}) should be carefully estimated because the divergent terms in (\ref{n2}) and (\ref{n2}) may occur finite and will not be mutually eliminated if ${\rm Re}\varepsilon_{p \pm} \neq 0$ in the presence of external field that may imply request for production of other sorts of charged particles. Of course, for an electrically neutral
system (for example, a mixture of positive-charged tachyons and electrons),
this effect may play visible role only beyond the classical or mean-field level when
the quantum exchange and correlation corrections to interaction are taken into
account. This question requires development in the further research.

The author is grateful to Konstantin Stepanyantz for inspiring discussions.

\newpage 

\end{document}